\documentclass[a4paper,11pt]{article}
\usepackage{jheppub} 
\usepackage{lineno}

\usepackage[dvipsnames]{xcolor}

\usepackage{enumerate}

\arxivnumber{2503.10565} 

\title{\boldmath Apparent teleportation of indistinguishable particles}

\author[a]{Marek Gazdzicki}
\author[b]{Daniel Kikola}
\author[a]{Ivan Pidhurskyi}
\author[a]{Leonardo Tinti}

\affiliation[a]{Jan Kochanowski University in Kielce,\\
Kielce, Poland}
\affiliation[b]{Warsaw University of Technology,\\
Warsaw, Poland}

\emailAdd{marek.gazdzicki@cern.ch}

\abstract{
Teleportation, introduced in science fiction literature, is an instantaneous change of the position of a macroscopic object. 
Two teleportation-like phenomena have been predicted by quantum mechanics: quantum teleportation and, more recently, quantum particle teleportation. 

Here, we introduce the third teleportation-like phenomenon—apparent teleportation. It seems to be a natural consequence of the Standard Model's indistinguishable elementary particles and antiparticles. We illustrate the idea within a 1+1D toy model of particle-antiparticle creation and space-time evolution obeying transport locality. Furthermore, we propose a novel method to observe apparent teleportation driven by strong interactions through measurements of correlations between the momenta of charm and anticharm hadrons in nuclear collisions. Observing the apparent teleportation would uncover the basic transport properties of indistinguishable particles.
}

\begin{document}
\maketitle
\flushbottom

\section{Introduction}
\label{sec:intro}

 \textbf{Teleportation}~\cite{Kaku2008} is rooted in science fiction.
    It describes the ability of macroscopic objects, such as humans, to reappear at a distant location without traversing space. Thus, the speed of light does not restrict the possibility of changing our location in space. Needless to say, 
    this also concerns signals. Details depend on the science fiction setup.
    
    \textbf{Quantum teleportation}~\cite{Bennett:1992tv} is the standard quantum physics phenomenon - transferring quantum-state properties over arbitrarily large distances in no time.
    In quantum teleportation, conditional probabilities of measurement outcomes reappear at a distant location using the entanglement of two particles. There is neither a super-luminal transfer of conserved quantities nor super-luminal signalling. The quantum teleportation is measured in numerous experiments~\mbox{\cite{Bouwmeester:1997slj, Yin:2017ips,Pfaff:2014lgp, Pirandola:2015nwo}.} 

    \textbf{Quantum particle teleportation}~\cite{Porras:2023fgn} is an instantaneous transfer of a single particle over a large distance predicted by standard quantum mechanics. The phenomenon appears for particles in a potential. It is induced by 
    frequently monitoring if the particle is at rest    
    (a peculiar quantum Zeno dynamics~\cite{Facchi:2000bs,Facchi:2002jgu,Facchi:2008nrb}).
    Quantum particle teleportation exhibits the properties of "science fiction" teleportation, allowing for the reappearance of conserved quantities and signalling. It was predicted in 2023 and has yet to be experimentally tested.

    \textbf{Apparent teleportation} (AT) is a concept introduced here. It is defined as the time evolution of a system of indistinguishable particles that cannot be mimicked by distinguishable particles moving with luminal or subluminal velocities. Of course, 
    when referring to distinguishable particles mimicking indistinguishable ones, we assume the observer is blind to particle labels.
    The apparent teleportation seemingly requires superluminal velocities of particles. This illusion arises from the fact that particle trajectories and velocities are defined for distinguishable particles but are undefined for indistinguishable ones.

\section{Results}
\label{sec:results}

Elementary particles and antiparticles of the Standard Model are indistinguishable bosons and fermions. Objects composed of many indistinguishable elementary particles and antiparticles are close to unique and thus can be approximately treated as distinguishable ones~\footnote{
An internally consistent model covering distinguishable and indistinguishable particles and antiparticles has to give identical predictions for a single particle or antiparticle in the system when treating it as distinguishable and indistinguishable.}. 
Classical physics is rooted in our daily experience within the world of macroscopic, distinguishable objects, including humans. In particular, it is well established that they can move only with velocities smaller than the speed of light.
However, applying time evolution limits of distinguishable objects to indistinguishable particles and antiparticles imposes human-centred experience on a fundamentally different world. The inadequacy is obvious if these limits are given in quantities that are undefined for indistinguishable particles and antiparticles, such as particle velocity. In this sense, apparent teleportation is the natural property of models postulating the indistinguishability of elementary particles and antiparticles.

In the following, we illustrate the concept of apparent teleportation within a toy model of particle-antiparticle creation, annihilation, and transport. Importantly, we propose a novel method which allows for observing the apparent teleportation in collisions of two nuclei, if it exists.

\subsection{Toy model of apparent teleportation}
\label{sec:cm}
    \textbf{Cell Model} is a dynamical model based on the 1+1D discrete-time Markov chain framework~\cite{Gazdzicki:2017rfe, Gazdzicki:2022zej}. Space is assumed to be a vector of $V$ discrete cells $(v_1, v_2, \ldots, v_V)$ arranged in 
    a ring. At the given time $t$, the system microstate is fully defined by a cell distribution of particles and antiparticles.
    The system's evolution in time is assumed to be discrete, and the
    time steps are numbered by $t$. During evolution, transitions occur between microstates.
    The transition probability from a microstate $X$ at $t$ to a 
    microstate $Y$ at $t+1$ depends only on the microstate $X$ - the basic assumption of Markov chains.  
    
    To illustrate the concept of apparent teleportation, we consider three types of reactions changing the system's microstates:
    \begin{itemize}
    \renewcommand{\labelitemi}{-}    
    \item
    particle-antiparticle creation in a single cell, 
    \item 
    particle-antiparticle annihilation in a single cell and,
    \item
    redistribution of particles and antiparticles between cells.
    \end{itemize}
    The first two reactions change the number of particles and antiparticles, whereas the third only changes the distribution of (anti) particles in cells.
    The difference between particle and antiparticle numbers of a given type is conserved in the whole system. This is the only conservation law that applies to the system.

    To summarise, the distribution of particles and antiparticles in cells generally differs at different time steps. This can be due to particle-antiparticle creation-annihilation reactions and particle and antiparticle redistribution.
    Now, we postulate that only changes obeying the transport locality are allowed during the redistribution.
    For distinguishable particles, the transport-locality requirement reduces to a requirement of each particle moving by no more than $\Delta$ cells.  In physics, this corresponds to
    particle velocities, which are limited by the speed of light in a vacuum. 
    
    For indistinguishable particles, the particles' trajectories and thus
    velocities are undefined. Thus, the transport locality condition for distinguishable particles does not apply to indistinguishable ones. The condition for indistinguishable particles and antiparticles is unknown, and experiments are needed to uncover it.
    
Here, we assume that the transport locality condition introduced for conserved particle number in Ref.~\cite{Gazdzicki:2022zej} (see below) is valid separately for redistributions of indistinguishable particles and antiparticles. 
It implies the following.
During a single time step, the particle (antiparticle) number in any interval of cells cannot be transported beyond an interval by $\Delta$ cells longer on the left and right. Moreover, it cannot be squeezed into an interval by $\Delta$ cells shorter on the left and right. The condition means that only those redistributions of particles (antiparticles) that can be mimicked by distinguishable particles (antiparticles) are possible.
The condition is given by two transport-locality inequalities~\cite{Gazdzicki:2022zej}:
\begin{align}
\label{eq:local_leq}
    \nonumber
    \sum_{j=i}^{i+k}n_j^X \leq \sum_{l=i-\Delta}^{i+k+\Delta} n_l^Y~, \\ 
    \sum_{l=i}^{i+k}n_l^Y \leq \sum_{j=i-\Delta}^{i+k+\Delta} n_j^X~,
\end{align}
where $k = 0, 1, \ldots $ and $n_j^X$, $n_l^Y$ are particle numbers in cells $j$, $l$ of $X$ and $Y$ microstates, respectively. If the (anti) particle number in a single cell is unlimited, the inequalities concern bosons; if the number is limited to 0 and 1, they concern fermions.

Concerning quantum mechanics, the Cell Model for indistinguishable particles mimics the time evolution of a coherent quantum state of particles on a one-dimensional space lattice. Not considering particles' momenta and initially allowing for all transitions corresponds to having the wave functions of (anti)particles delocalise in the whole system. 
The Heisenberg uncertainty principle provides a simple and intuitive lower bound on momentum, which relates directly to quantum de-localisation and coherence:
$ \Delta p \approx \hbar/ (2 L)$, where $L$ is the system size.   
The time steps correspond to the time intervals between measurements of (anti) particle positions. Out of all possible transitions, the transport-locality condition (\ref{eq:local_leq}) allows only those that obey the speed-of-light limit. This, together with the particle-antiparticle creation and annihilation processes, provides the model with the basic properties of quantum mechanics and relativistic physics.

\begin{figure}[ht]
\includegraphics[width=0.90\textwidth]{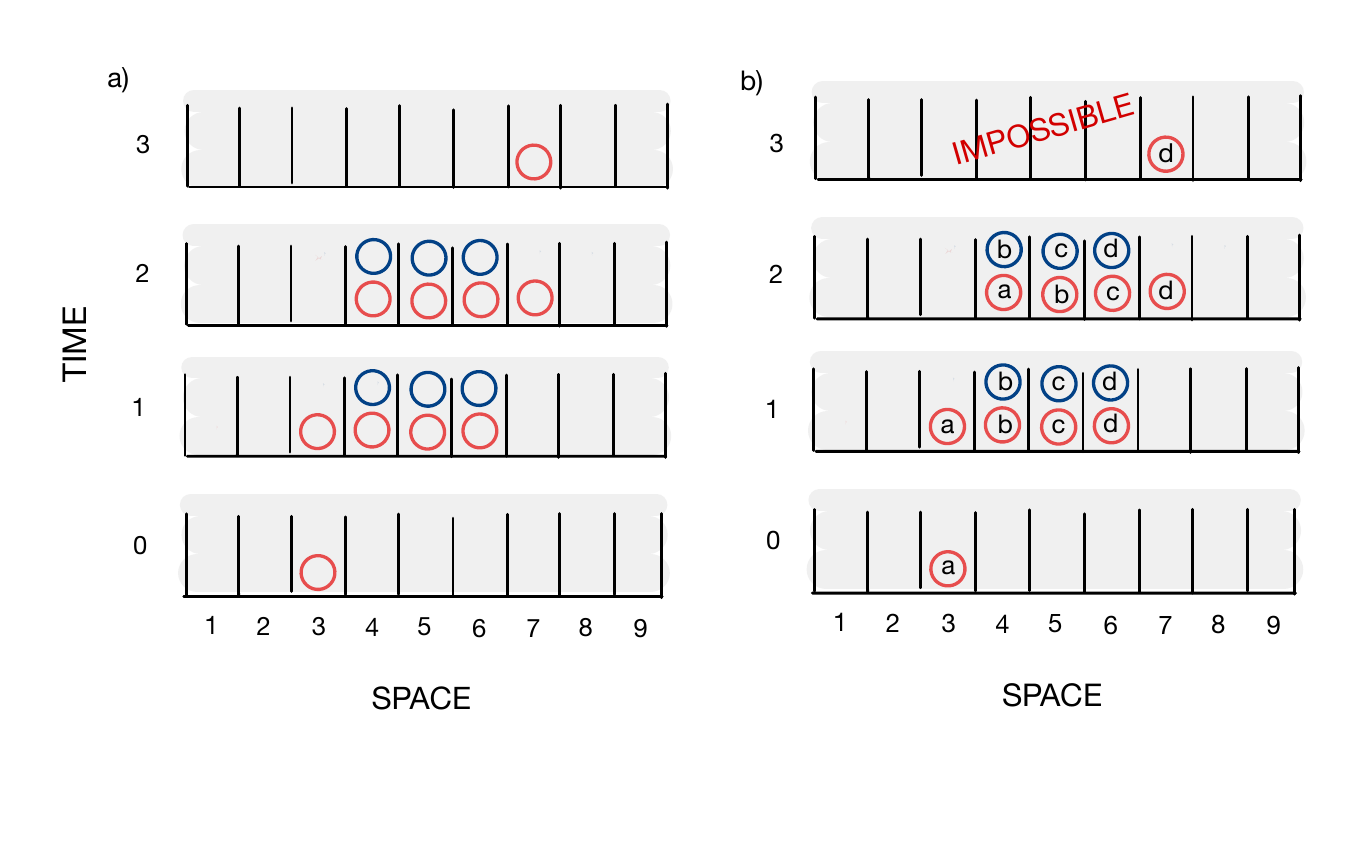} 
\caption{
The simple setup illustrates the apparent teleportation of indistinguishable particles within the Cell Model. The maximum distance a distinguishable particle and antiparticle can move during a time step is set to $\Delta = 1$.
Three types of transitions — pair creation, local transport, and pair annihilation — are possible. 
\textbf{a} The example process with indistinguishable particles starting from a single particle in cell three at $t=0$ and ending in a single particle in cell seven at $t=3$.
\textbf{b} Unsuccessful attempt to mimic the process by distinguishable particles. Three pairs of particles and antiparticles with identical labels appear at $t=1$. Particles are shifted by one cell to the right at $t =2$. 
In particular, the particle "d" appears in cell seven.  
The particle number distributions for indistinguishable and distinguishable particles are identical at $t = 0$, 1 and 2.
The pair annihilation reactions in cells 4, 5 and 6 are impossible during the third step because cells contain particles and antiparticles with different labels. For example, cell 4 contains
particle "a" and antiparticle "b" at $t =2$. See the text for more details.
}
\label{fig:tele}
\end{figure}

\textbf{Apparent teleportation within Cell Model} is illustrated in Fig.~\ref{fig:tele}. An example of the system's time evolution with $V = 9$ cells and periodic boundary conditions is shown. 
Cells contain "background" particles that carry total energy, depicted by the grey band. Particles and antiparticles of interest are depicted by red and blue circles, respectively. 
For simplicity, we assume that the speed-of-light limit implies $\Delta = 1$. The difference between the total number of particles and antiparticles is assumed to be conserved. 
    
The case of indistinguishable particles and antiparticles is presented in the  plot~\textbf{a} of Fig.~\ref{fig:tele}. We postulate that
only one particle exists at $t=0$, and it is in cell 3. Three pairs of particles and antiparticles appear in cells 4, 5 and 6 at $t=1$. 
At $t=2$, the particle number distribution is shifted by one cell to the right by the transport-local transition. This is allowed 
by the transport-locality inequalities~(\ref{eq:local_leq}).
Then, at $t=3$, particle-antiparticle pairs annihilate, leaving a single particle in cell seven.      
    
The plot~\textbf{b} of Fig.~\ref{fig:tele} shows an unsuccessful attempt to mimic the above process using distinguishable particles.
In the example presented here, they are labelled by the letters "a", "b", "c", and "d". Similarly, antiparticles have unique labels. 
Only creating a particle and an antiparticle with identical labels is possible. Similarly, only a particle and an antiparticle with identical labels can annihilate. 
    
For the label-blind observer, the particle and antiparticle number distributions of indistinguishable and distinguishable particles are identical at $t = 0, 1$, and 2. But the similarity breaks at $t=3$. This is because annihilating particles and antiparticles with different labels is impossible.
Of course, distinguishable particles can mimic the process depicted in 
the plot~\textbf{a} if the transport-locality requirement is lifted. Particle "a" can be "teleported" by four cells in three steps from cell $3$ to $7$.

The example construction of the process shown in Fig.~\ref{fig:tele} can be easily generalised, allowing for an arbitrary redistribution of indistinguishable boson-like 
particles and antiparticles in three time steps.
This includes apparent teleportations - the processes which 
seemingly violate the speed of light limit. 

We note that the microscopic mechanism of the apparent teleportation illustrated in Fig.~\ref{fig:tele}~\textbf{a} cannot be used for superluminal signalling. An observer in cell seven detecting the particle
at $t = 3$ cannot be sure it is caused by injecting a particle at $t=0$ in cell three. For example, this can occur when a particle and an antiparticle pair are created in cell six independently of events in cell three. Then a particle appears in cell seven.

The probability of a given process depends on the dynamics which in the Cell Model is encapsulated within the transition matrix. One can speculate that at sufficiently high energy densities, the transition matrix of indistinguishable particles approaches maximum symmetry (the microstate symmetry~\cite{Gazdzicki:2022zej}), and all processes have the same probability. This would imply that the system is ``born in equilibrium"~\cite{Hagedorn:1965st} - the probability of any microstate appearing at any time is equal.
This may explain the puzzle of fast equilibration in heavy-ion collisions at high energies~\cite{Schlichting:2019abc}.
The apparent teleportation also may explain the quark-gluon paradox formulated in Ref.~\cite{Miskowiec:2007qv}.
The transition of a large volume of quark-gluon plasma created in collisions of two atomic nuclei (see below for details) to colourless hadrons appears to require superluminal transport of colour charge to ensure local colour neutrality. Quarks and gluons are indistinguishable particles.
Thus, the apparent teleportation may also serve as the explanation of the quark-gluon plasma paradox.

The above discussion motivates the question of whether the apparent teleportation, if it exists,
can be observed experimentally. This question is addressed in the following subsection.

\vspace{0.5 cm}
\subsection{Observing apparent teleportation}

\textbf{System of quarks and gluons} created in nucleus-nucleus collisions at high energies is the highest energy density system 
($\epsilon > 1$~GeV/fm$^3$ $\approx 1.6 \cdot 10^{35}$~J/m$^3$)   
created under conditions controlled in the laboratory~\cite{Florkowski:Phenomenology}. 
Its equilibrium state is called quark-gluon plasma (QGP)~\cite{Shuryak:1980tp}. The plasma expands and cools down, and the transition to colour-neutral hadrons - small bags of quarks and gluons -occurs at the hadronisation (transition) temperature $T_{\textrm{HAD}} \approx 150$~MeV~\cite{Aoki:2006br}.

Six types of quarks (fermions) $u$, $d$, $s$, $c$, $t$ and $b$ and the corresponding antiquarks and the eight types of gluons (bosons) can be created during the collision. 
Massless gluons and light $u$ and $d$ quarks and antiquarks are the most abundant. Production of heavier quarks is suppressed even at the top collision energies of the CERN LHC~\cite{ALICE:2013mez}. 
The threshold for QGP creation in heavy-ion collisions is located at $\sqrt{s_{NN}} \approx 10$~GeV~\cite{NA49:2002pzu, NA49:2007stj}. For a review, see also Refs.~\cite{Gazdzicki:2010iv, Andronov:2022cna}.
The fixed target experiments at the CERN SPS measure heavy-ion collisions in the energy range $\sqrt{s_{NN}} \approx 5 - 20$~GeV. At the top SPS energy, just above the QGP threshold energy, the mean number of light quarks and gluons produced in central collisions of two lead nuclei is on the order of 1000~\cite{Gazdzicki:1998vd}.
The corresponding numbers for strange and charm quarks are $\approx$~100~\cite{Gazdzicki:1998vd} and $\approx$~1~\cite{Merzlaya:2024cbt}, respectively.

The local creation and annihilation of $c$ and $\bar{c}$ pairs implies that the energy and momentum scale of the interaction determines the $c$ and $\bar{c}$ space-time separation via the uncertainty principle.
The scale ranges between:
\begin{itemize}
\renewcommand{\labelitemi}{-}
\item
the hard-scattering scale of the isolated reaction,
$\hbar c/(2 m_c) \approx 0.1$~fm, with $m_c \approx$~1.3~GeV being the charm quark mass, and
\item 
the thermal scale of the reaction in the dense quark-gluon plasma,
$\hbar c/T \approx 1$~fm, with $T \approx$~0.2~GeV being the quark-gluon plasma temperature just above the transition (hadronisation) temperature.
\end{itemize}

The spatial extent and lifetime of the system created in central Pb+Pb collisions are on the order of approximately 10 fm, the diameter of the lead nucleus.
 It is much larger than the space-time cell in which local creation and annihilation of $c$ and $\bar{c}$ quarks occur. It opens the possibility of observing the apparent teleportation in the collisions.

\vspace{0.5 cm}
\textbf{Observing apparent teleportation in heavy-ion collisions} is likely to be a challenging task. 
This is due to two reasons:
\begin{itemize}
\renewcommand{\labelitemi}{-}
    \item The experimental challenge:
    We cannot measure the location in the space of a single charm quark created in a collision at two different times.
    Hence, the simple setup presented in Fig.~\ref{fig:tele} cannot be used
    for observing the apparent teleportation.
    \item The theoretical challenge: Quantum Chromodynamics, the theory of strong interactions, does not provide quantitative predictions for multi-particle correlations in heavy-ion collisions; for further discussion, 
    see Ref.~\cite{Gazdzicki:2023niq}.
\end{itemize}
However, with the help of well-established heavy-ion models, it may be possible to provide experimental evidence of the apparent teleportation by measuring momentum correlations between $c$ and $\bar{c}$ quarks.
The idea is rooted in our previous paper~\cite{Gazdzicki:2023niq}, which suggests the use of momentum correlations of charm-anticharm hadrons to extract information on spatial correlations of $c$ and $\bar{c}$ quarks at hadronisation. 
The paper reviews the experimental results on charm hadron correlations and collective flow. It also discusses the importance of having collisions with no more than one $c\bar{c}$ pair created. 
We note that the requirement of a single $c\bar{c}$ pair in a collision can be reconciled with the toy mechanism of the apparent teleportation presented 
in Sec.~\ref{sec:cm}, assuming the coherent creation and annihilation of $c\bar{c}$ pairs in a short time interval.\\

\begin{figure}[ht!]
\centering
\includegraphics[width=0.9\textwidth]{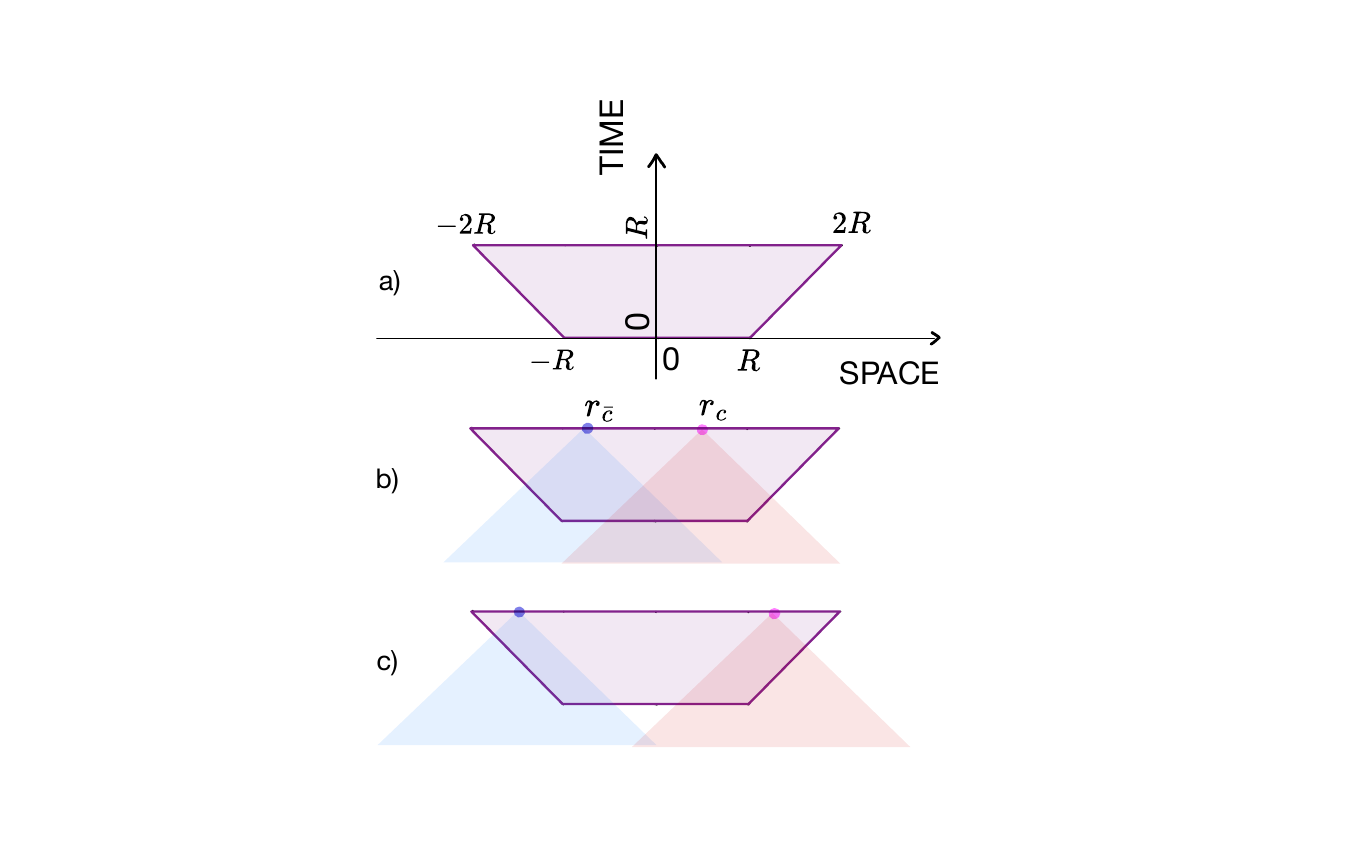} 
\caption{
The simplest illustration of observing the apparent teleportation in heavy-ion collisions at high energies using a single pair of $c$ and $\bar{c}$ quarks.
The plot~\textbf{a} shows the space-time (1+1D) evolution of the quark-gluon plasma created at $t=0$ in the spatial region from $-R$ to $R$. The hadronisation is assumed to occur at $t=R$ when the system extends from $-2R$ to $2R$. 
The plot~\textbf{b} shows an example of the positions of the $c$ and $\bar{c}$ quarks at hadronisation, denoted as $r_c$ and $r_{\bar{c}}$, respectively. In this example, the past light cones of the quarks overlap within the collision light cone.
This overlap is required for the subluminal movement of distinguishable $c$ and $\bar{c}$ quarks created at a common space-time point. We refer to this as the local-distinguishable
condition.
The plot~\textbf{c} depicts the positions of the $c$ and $\bar{c}$ quarks, which do not obey the condition and can only be explained by the apparent teleportation.
}
\label{fig:HIC}
\end{figure} 

The standard model of heavy-ion collisions at high energies describes the collision as follows. 
\begin{enumerate}[(a)]
\item 
Initial stage: a high-density quark-gluon plasma is created. Theoretical approaches addressing this stage are discussed in Ref.~\cite{Schlichting:2019abc}.
\item
Expansion stage: the plasma expands~\cite{Landau:1953gs}, cooling down to the hadronisation temperature ($T_{\textrm{HAD}} \approx 150$~MeV in the local frame of the flowing matter). This stage is modelled using relativistic hydrodynamics~\cite{Florkowski:Phenomenology}.
\item
Hadronisation stage: the plasma is converted into hadrons and resonances following statistical rules~\cite{Hagedorn:1980kb, Becattini:2005xt} applied in the rest frame of a plasma fluid element. As a result, 
hadron momenta are determined by their flow acquired during the expansion stage and by the hadronisation (a local statistical process).
\item
Free-streaming stage: resonances decay, and non-interacting hadrons freely stream in a vacuum to a detector.
\end{enumerate}

We illustrate the idea of observing the apparent teleportation with the help of the schematic 1+1D model of the collisions sketched in Fig.~\ref{fig:HIC}.
For simplicity, we assume that the initial stage is created at $t = 0$ and extends in space from $-R$ to $R$. The hadronisation occurs at $t = R$; at this stage, the system extends from $-2R$ to $2R$. We also neglect the finite size of the space-time cell of the local $c\bar{c}$ pair creation.
In the plots~\textbf{b} and \textbf{c}, two examples of the hadronisation points of a single $c\bar{c}$ pair created in the collision are depicted as $r_c$ and $r_{\bar{c}}$. The plots also show the past light cones of the charm and anticharm quarks (represented by the blue and red areas).

Let us assume the $c$ and $\bar{c}$ quarks were created in the same space-time point within the collision light cone and moved to hadronisation as distinguishable particles with subluminal velocities.
In this case, the creation point has to be located inside the overlap between the collision light cone and the past-light cones of $c$ and $\bar{c}$ quarks - the local-distinguishable (LD) condition.
The $c\bar{c}$ pair depicted in Fig.~\ref{fig:HIC}~\textbf{b} fulfils the requirement, but the pair depicted in the plot~\textbf{c} does not.
The latter implies that the process was not local-distinguishable, and one needs the apparent teleportation to explain it.

Within the example shown in Fig.~\ref{fig:HIC}, the LD condition reads $| r_c - r_{\bar{c}} | \leq 2R$.
The apparent teleportation is required if the LD condition is violated, $| r_c - r_{\bar{c}} | > 2R$.
The positions of quarks at hadronisation are not measured. Instead, experiments measure velocities (momenta) of charm hadrons in the free-streaming stage.
We relate them in two steps: first, we introduce the correlation between the quark position and its flow velocity, and second, we unfold the flow velocity smearing due to the quark interactions with the medium and hadronisation.

\begin{figure}[ht!]
\includegraphics[width=0.9\textwidth]{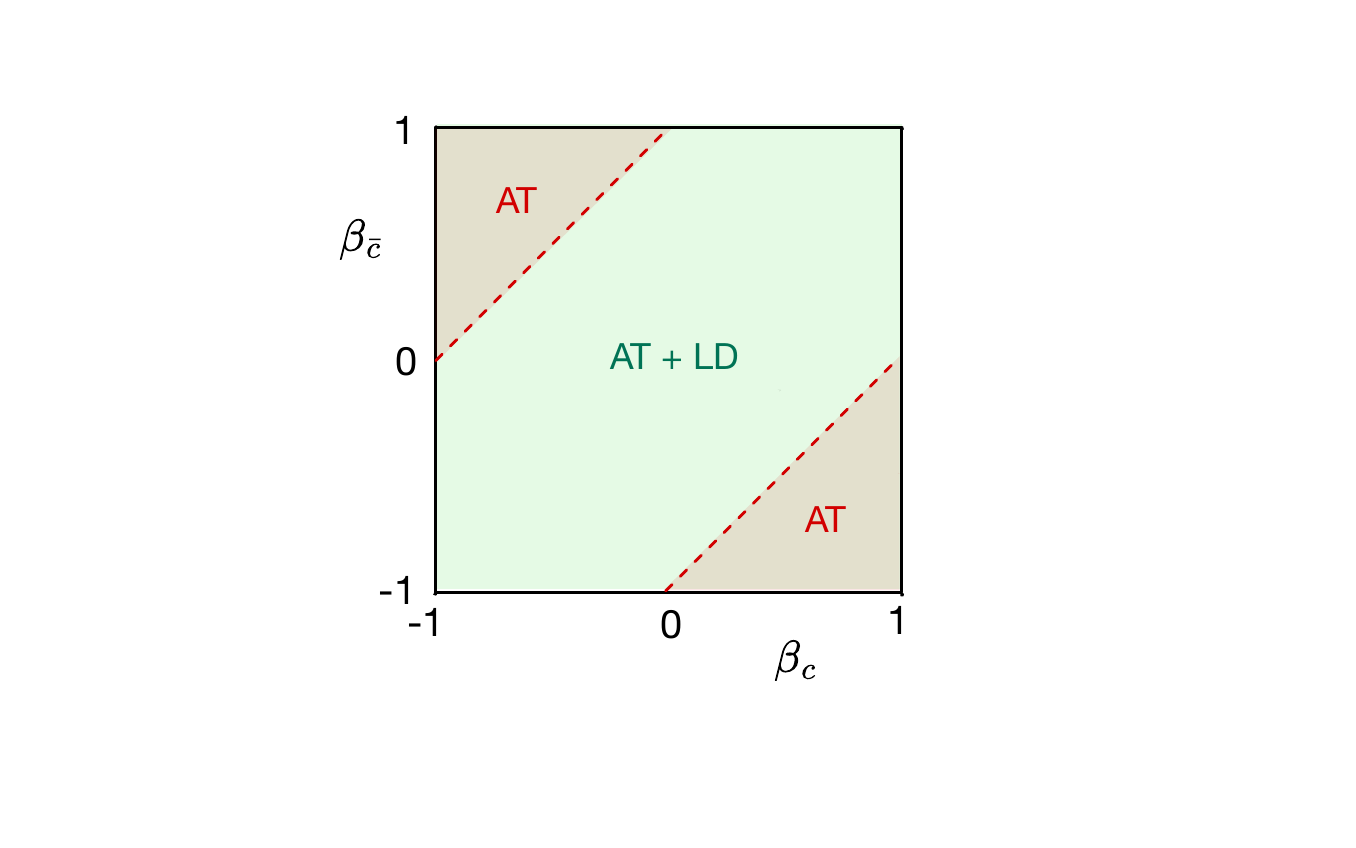} 
\vspace*{-1cm}
\caption{
The simplest 1+1D model of heavy-ion collisions and charm production explains the concept of observing apparent teleportation.
Processes resulting in $c$ and $\bar{c}$ quarks at hadronisation with flow velocities $\beta_c$ and $\beta_{\bar{c}}$ are depicted as AT - apparent teleportation and LD - local distinguishable.
The corners cut by the red dashed lines are forbidden for the local distinguishable but allowed for the apparent teleportation processes.
}
\label{fig:Exlcusion}
\end{figure}

Let us assume that the velocities of charm quarks at hadronisation are equal to the flow velocities of matter at the hadronisation points. For simplicity, we also assume that the flow velocity is proportional to the distance from the collision centre. From these assumptions, it follows that the apparent teleportation is required if $| \beta_c - \beta_{\bar{c}} | > 1$, where $\beta_c$ and $\beta_{\bar{c}}$ are $c$ and $\bar{c}$ quark velocities scaled by the light velocity. The pairs which obey the complementary condition,
$| \beta_c - \beta_{\bar{c}} | \leq 1$, may come either from LD or AT processes.
Allowed processes leading to pairs with given $\beta_c$ and $\beta_{\bar{c}}$ values are depicted in Fig.~\ref{fig:Exlcusion}.

After the hadronisation stage, the carriers of the \( c \) and \( \bar{c} \) quarks are charm and anticharm hadrons. For simplicity, we assume these are \( D \) and \( \bar{D} \) mesons.
The free-streaming velocities of these mesons, \( \beta_D \) and \( \beta_{\bar{D}} \), are superpositions of the quark flow velocities ( \( \beta_c \), \( \beta_{\bar{c}} \) ) and a stochastic, non-flow component. The non-flow component may arise from initial momenta of quarks, their diffusion in quark-gluon plasma, hadronisation, and the following decay of resonances. 
In Methods~A, we present a didactic illustration of the effect of the smearing of the flow components by the stochastic uncorrelated processes.
We show the results obtained within the schematic 1+1D model with the Hubble-like initial and hadronisation hypersurfaces. 

Finally, in Methods~B, the 1+3D model of high-energy nucleus-nucleus collisions, with properties that allow for the prediction of experimentally measured quantities, is presented. It utilises the Blast-Wave model~\cite{Schnedermann:1993ws} of radial flow.
The results indicate that the distribution of the opening angle between 
momenta of \( D \) and \( \bar{D} \) mesons may yield evidence for the apparent teleportation.
Assuming that the stochastic components of \( D \) and \( \bar{D} \) are uncorrelated, this distribution can be corrected for the smearing in an almost model-independent manner. The procedure is detailed in Ref.~\cite{Gazdzicki:2023niq}.

The initial \( c \) and \( \bar{c} \) momenta may be correlated due to energy-momentum conservation. Remnants of this correlation might persist throughout the transport process, leading to correlated non-flow components in the \( D \) and \( \bar{D} \) momenta. 
At collision energies where the creation of \( c \) and \( \bar{c} \) quarks occurs near the threshold—such that no more than one \( c\bar{c} \) pair is produced per collision—it is reasonable to expect that the initial quark momenta are small. Consequently, the correlated non-flow component is likely to be negligibly small.
A quantitative estimate of this effect lies beyond the scope of the present paper.

The apparent teleportation may render all transitions possible. 
One may speculate that at very high energy densities, the microstate-symmetric transition matrix~\cite{Gazdzicki:2022zej} may govern the system's evolution. 
This implies that \( c \) and \( \bar{c} \) quarks, and consequently \( D \) and \( \bar{D} \) hadrons, are produced independently in space-time and momentum.
The joint probability density function \( \rho(\beta_D, \beta_{\bar{D}}) \) factorises as:
\begin{equation}
    \rho(\beta_D, \beta_{\bar{D}}) = \rho(\beta_D) \cdot \rho(\beta_{\bar{D}})~,
\label{eq:factorization_1D}
\end{equation}
and, in the general 1+3D case:
\begin{equation}
    \rho(\mathbf{p}_D, \mathbf{p}_{\bar{D}}) = \rho(\mathbf{p}_D) \cdot \rho(\mathbf{p}_{\bar{D}})~,
\label{eq:factorization_3D}
\end{equation}
where \( \mathbf{p}_D \) and \( \mathbf{p}_{\bar{D}} \) are the momentum vectors of 
\( D \) and \( \bar{D} \) hadrons, respectively.
Contemporary experiments can test this extreme prediction following the apparent teleportation hypothesis soon~\cite{Gazdzicki:2023niq}. This requires measurements of sufficiently high statistics of charm and anticharm hadron pairs
produced in individual collisions with a mean multiplicity of pairs 
below one~\cite{Gazdzicki:2023niq}.

\section{Discussion}
\label{sec:discussion}

We argue that the apparent teleportation of indistinguishable particles naturally follows from the Standard Model postulate of elementary particles being indistinguishable. Thus, apparent teleportation is expected in many systems of indistinguishable particles and antiparticles.

The presented toy-model example of the apparent teleportation mechanism suggests that the coherent creation and annihilation of particle-antiparticle pairs is required for apparent teleportation. Thus, apparent teleportation is expected to be more popular in high-energy-density systems. 
It makes it difficult to observe experimentally.
This is because the high-density systems created in laboratories typically have small spatial and temporal dimensions. Therefore, precisely measuring particle number distribution in short time intervals may be challenging. 

We provide an example of overcoming the problem of observing the apparent teleportation. It can be done by measuring
momentum correlation of charm and anticharm hadrons produced in collisions of two sufficiently heavy atomic nuclei, with only one pair of charm and anticharm quarks created.
Contemporary experiments can perform the measurement; however, interpreting results on apparent teleportation will depend on modelling the time evolution of the strongly interacting system.

To strengthen the conclusions on the apparent teleportation, other experimental setups should be considered. The example presented above concerns apparent teleportation driven by strong interactions. It can be straightforwardly extended from the charm-anticharm hadron correlations to the bottom-antibottom hadron correlations. The mean number of $b$ and $\bar{b}$ pairs, $\langle b\bar{b} \rangle$, produced in central Pb+Pb collisions at the top LHC energy is expected to be on the order of 10~\cite{He:2022tod}. Thus, by lowering the LHC beam energy, the condition $\langle b\bar{b} \rangle \approx 1$ can be reached. Alternatively, one can consider the possibility of the measurements at the top RHIC energy by the operational sPHENIX experiment~\cite{sPHENIX:HF}. The corresponding measurements of bottom-antibottom hadron correlations may be a challenging task due to the low detection probability for central Pb+Pb collisions recorded in the collider mode.

The possibility of using strange-antistrange hadron correlations to search for the apparent teleportation should also be considered. The requirement of a mean number of strange and antistrange hadrons being on the order of one implies measuring their correlations in central Pb+Pb collisions at collision energies of several GeV~\cite{Blume:2011sb}. These are energies of the future CBM experiment at FAIR~\cite{CBM:2016kpk}. They are below the onset of the quark-gluon plasma creation~\cite{NA49:2002pzu,NA49:2007stj}. Thus, the carriers of strangeness are different species of strange and antistrange hadrons and resonances. This brings the system closer to the system of distinguishable particles for which, by definition, the apparent teleportation does not exist. Thus, one expects that at energies below the onset of quark-gluon plasma creation, the chance to observe the apparent teleportation is reduced.
This concerns the search for the apparent teleportation via correlations of strange-antistrange hadrons.

Finally, we consider the possibility of observing apparent teleportation driven by electromagnetic interactions. It seems that the simplest setup consists of an extended system of photons at energy density close to the Schwinger limit for spontaneous creation of 
$e^+ - e^-$ pairs~\cite{Schwinger:1951nm}. In terms of the electric field strength, the Schwinger limit is about 
$1.3 \cdot 10^{18}$~V/m, which corresponds to the energy density $\approx 10^{18}$~J/m$^3$
($\approx 10^6$~GeV/fm$^3$). Existing and planned ultra-intense laser facilities, like the Extreme Light Infrastructure in Europe,
are far below the Schwinger limit. 

Also, ultra-peripheral heavy-ion collisions (UPCs) at the LHC may offer an opportunity to search for apparent teleportation driven by electromagnetic interactions. In such collisions, a photon interacts with a strong electromagnetic field created by a heavy ion moving with ultra-relativistic velocities. These conditions facilitate the production of $c\bar{c}$ pairs; see, for example, measurements of charmonium production in the UPCs by the LHC experiments~\cite{CMS:2023snh, ALICE:2021gpt, LHCb:2021bfl} or a recent report on $D^0$ production in UPCs by the CMS experiment~\cite{CMS-PAS-HIN-24-003}. Thus, in principle, a measurement of $D - \bar{D}$ correlation is possible, simultaneously fulfilling the requirement of a single charm pair created in a collision. However, the probability of apparent teleportation is likely significantly lower under these conditions, and a feasibility study is necessary to determine an optimal experimental setup to observe it. Other probes (strangeness, $e^+ - e^-$ pairs) could also be considered.

If it exists, the apparent teleportation may explain two puzzles about strong interactions at high energy densities: the rapid equilibration of the created system of quarks and gluons and its rapid hadronisation, fulfilling the requirement of local colour neutrality.

\newpage

\appendix
\section{The 1+1D Hubble-like model}
\label{app:A}

Here, we illustrate the effect of the smearing by stochastic uncorrelated processes, taking as an example the statistical hadronisation of charm and anticharm quarks.
We present the results obtained within the schematic 1+1D model with the Hubble-like initial and hadronisation hypersurfaces. 

\begin{figure}[h!]
\centering
  \includegraphics[width=0.4\linewidth]{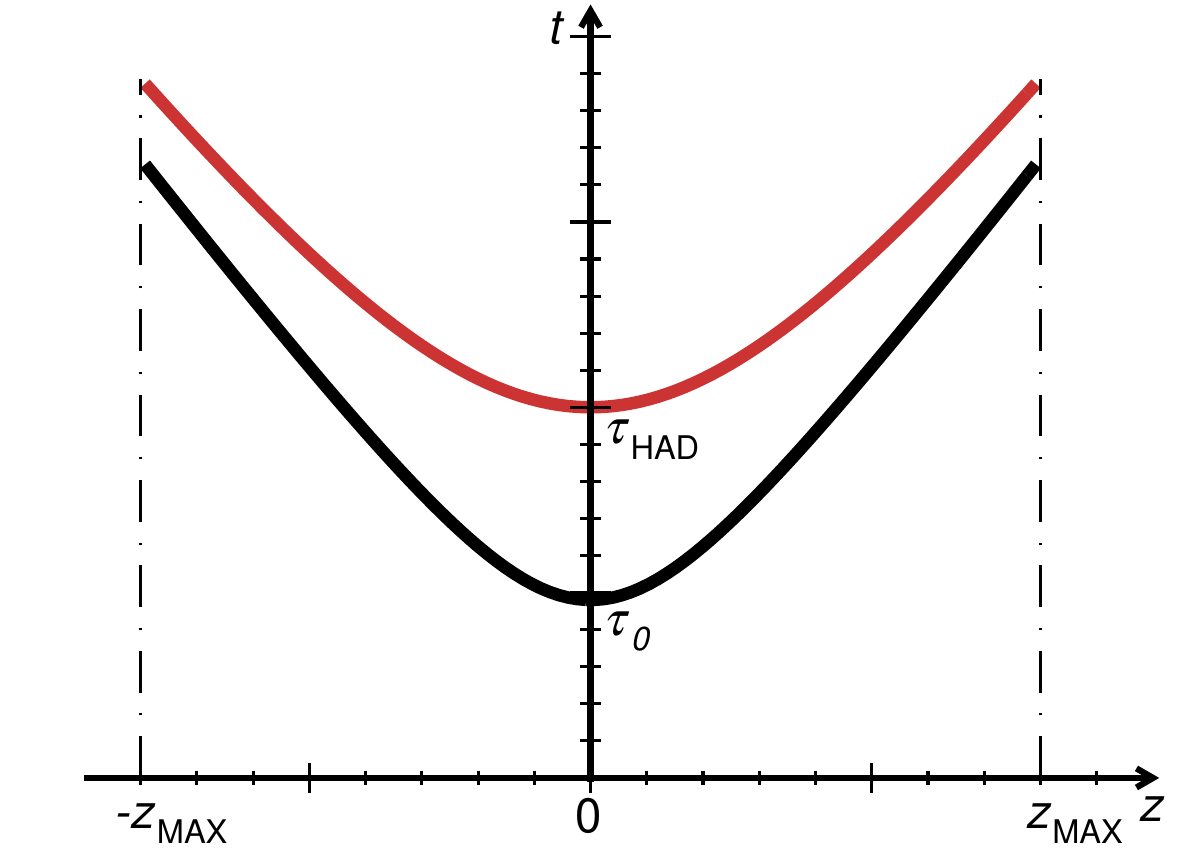}
\caption{
Sketch of the 1+1D Hubble-like model of the space-time evolution of the system created in lead-lead collisions. The black curve shows the initial hypersurface defined by the proper time $\tau_0$ at which the system is formed and its expansion starts. The red curve presents the hadronisation hypersurface given by the $\tau_{\mathrm{HAD}}$ parameter.}
\label{fig:sketch:double:Hubble}
\end{figure}

The shape of the initial system is described by its initial hypersurface:
\begin{equation}
\label{eq:initsur}
t^2 = z^2 + \tau_{0}^2~,
\end{equation}
where $\tau_{0}$ is the initial proper time the system is formed, and $z$ is the distance from the system's centre. 
The system then undergoes a Hubble-like expansion, and the hadronisation hypersurface is taken as
\begin{equation}
\label{eq:hadsur}
t^2 = z^2 + \tau_{\mathrm{HAD}}^2~, 
\end{equation}
where $\tau_{\mathrm{HAD}}$ is the hadronisation proper time. 
The expanding matter is assumed to be in local equilibrium corresponding to
temperatures $T_{0}$ and $T_{\mathrm{HAD}}$ at proper times $\tau_{0}$ and
$\tau_{\mathrm{HAD}}$, respectively. The matter density is assumed to be uniform along the hadronisation hypersurface, which is limited in $z$ as
$|z| < \tau_{\mathrm{HAD}} + R_{\mathrm{MAX}}$.
In the following the parameters $\tau_{\mathrm{HAD}}$ and $R_{\mathrm{MAX}}$
are fixed to 10~fm and 6~fm, being on the order of the Pb nucleus radius.
Figure~\ref{fig:sketch:double:Hubble} shows a sketch illustrating the hypersurfaces and indicating their parameters.

For the Hubble-like flow, the energy conservation relates the initial parameters $\tau_{0}$ and $T_{0}$ 
to the hadronisation parameters $\tau_{\mathrm{HAD}}$ and $T_{\mathrm{HAD}}$ as:
\begin{equation}
\label{eq:2H}
\frac{\tau_0}{\tau_{\mathrm{HAD}}} = 1 - \frac{3}{4}\;\ln \left( \frac{T_{0}}{T_{\mathrm{HAD}}}\right)~.
\end{equation}
The hadronisation temperature is set to the well-known value from the lattice QCD, 
$T_{\textrm{HAD}} =$~150~MeV~\cite{Aoki:2006br}. The initial temperature $T_0$ has to be larger than the hadronisation temperature. Below, we present examples calculated for $T_0 = 200$~MeV and $T_0 = 300$~MeV. The latter value is close to the initial temperature given by the Fermi-Landau initial conditions~\cite{Gazdzicki:1998vd} at the top energy CERN SPS. It can be considered as the upper limit of $T_0$ at SPS. The remaining parameter of the model $\tau_0$ is calculated using Eq.~\ref{eq:2H}.
The velocity of the fluid element and (anti) charm quarks at the hadronisation is given by the Hubble flow assumption, $u = z/\tau_{\mathrm{HAD}}$. 

At the hadronisation, (anti) charm quarks are converted into (anti) $D$ mesons. To model this process, the momentum of the (anti) charmed mesons at the fluid rest frame $p_{D}'$ is drawn from the statistical-like distribution: 
\begin{equation}
\label{eq:1DB}
dN/dp_{D}' ~~\propto~~ {p'_{D}}^2~\cdot~
e^{ -\sqrt{m^2_{D} + {p_{D}'}^2}~/~T_{\mathrm{HAD}}}~,
\end{equation}
where we use mass of $D^0$ meson $m_{D} = 1.869$~GeV. Finally, the momentum $p_D$ in the laboratory frame at hadronisation is calculated by boosting $p_D'$ with the fluid-element velocity. To assess how the statistical hadronisation influences the expected signal of the apparent teleportation, we also considered a case with $p_{D}' = 0$, thus removing the effect of statistical smearing of the momentum of (anti) charm mesons.

\begin{figure}
\centering
  \includegraphics[width=.45\linewidth]{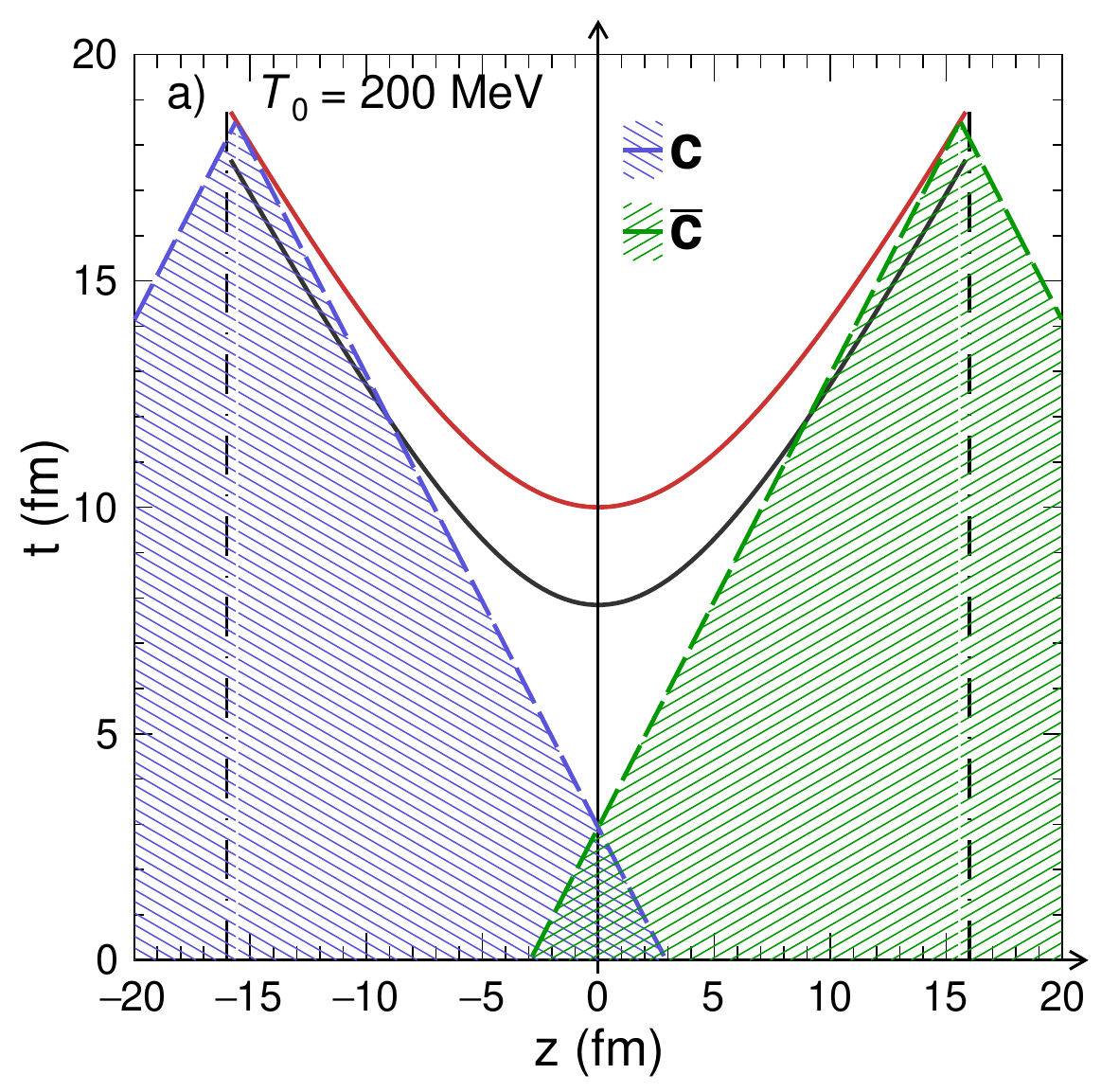}
  \includegraphics[width=.45\linewidth]{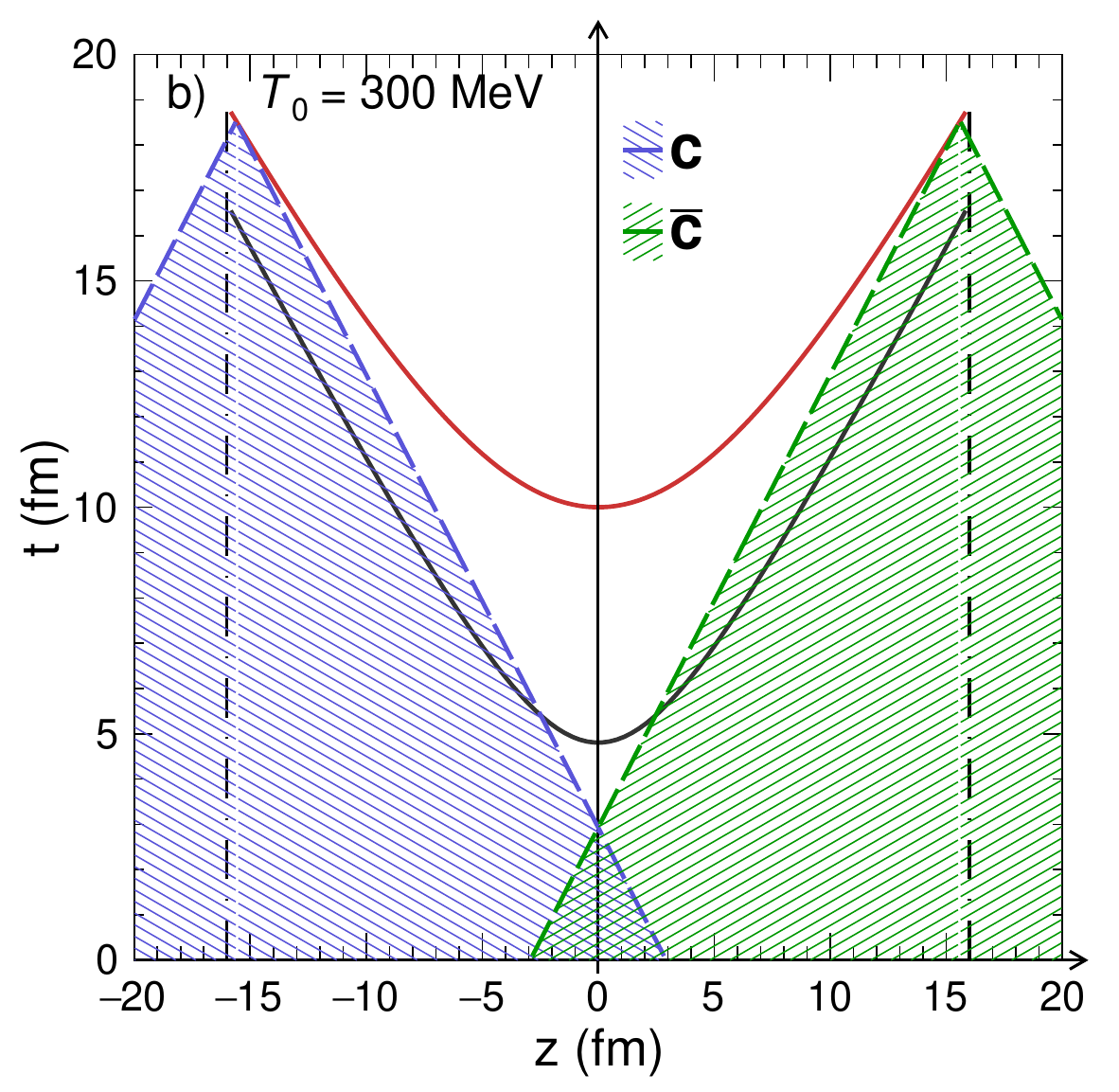}
\caption{
Examples of hadronisation points of $c$ and $\bar{c}$ quarks that can only
result from the apparent hadronisation. The initial temperature is set to $T_0 = 200$~MeV~(\textbf{a}) and $T_0 = 300$~MeV~(\textbf{b}). The black lines indicate the initial hypersurface, and the hadronisation hypersurface is shown by red lines.}
\label{fig:examples:teleport}
\end{figure}

Figure~\ref{fig:examples:teleport} shows two examples ($T_0 = 200$~MeV~(\textbf{a}) and $T_0 = 300$~MeV~(\textbf{b})) of the hadronisation points of charm and anti-charm quarks requiring the apparent teleportation - the past light cones of $c$ and $\bar{c}$ quarks overlap only below the initial hypersurface. We notice that the larger $T_0$ is, the smaller the space for separating the apparent teleportation from the local-distinguishable processes. The apparent teleportation processes can be separated up to $T_0 \approx 400$~MeV within the model, with its other parameters set to the values given in the text.

\begin{figure}
\centering
\includegraphics[width=.9\linewidth]{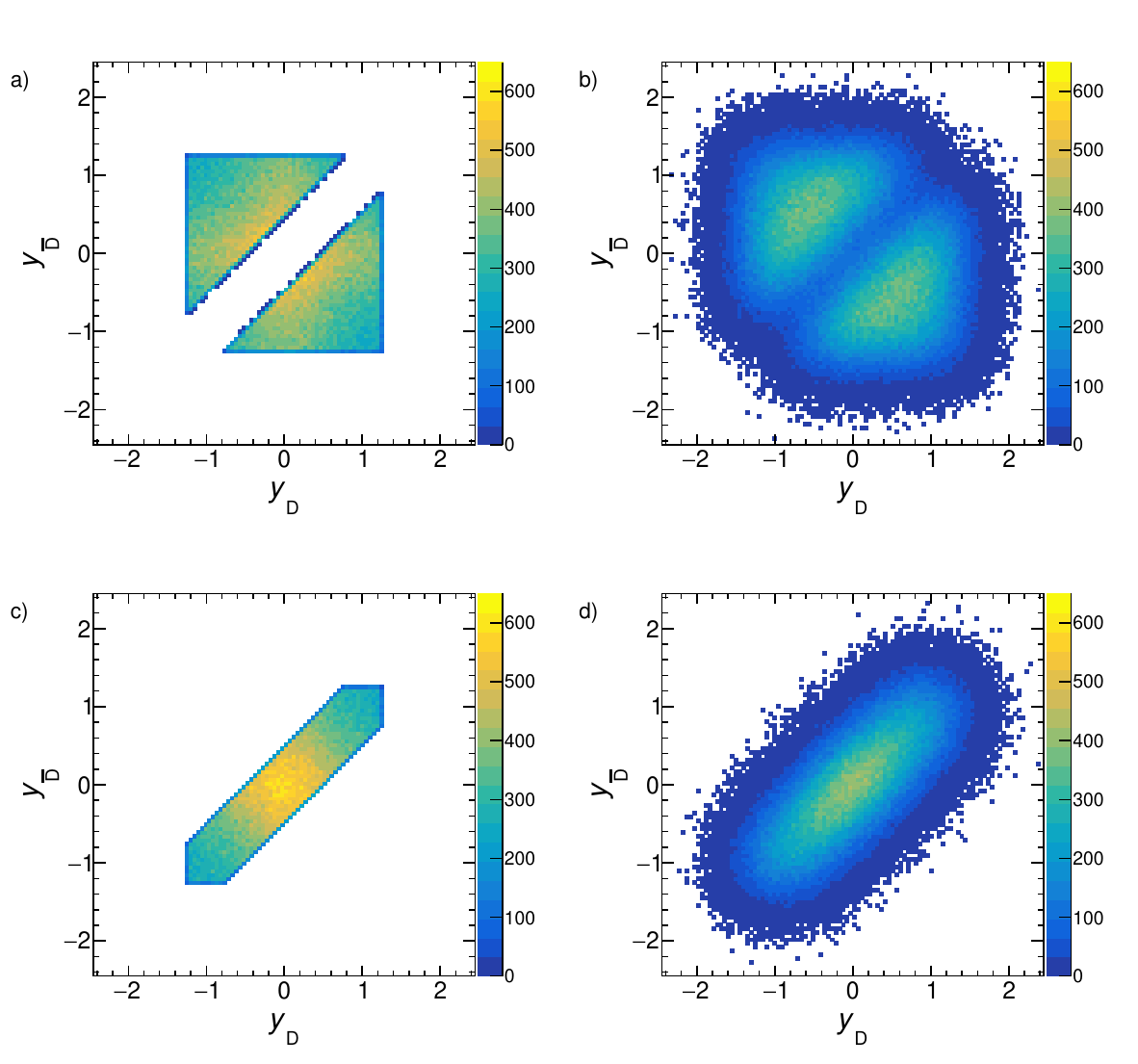}
\caption{
Joint rapidity distributions of $D$ and $\bar{D}$ mesons for the initial temperature $T_0 = 200$~MeV. 
The plots~\textbf{a} and~\textbf{b} show the distributions for $D$ and $\bar{D}$ pairs, which can result only from the apparent teleportation.
The plots~\textbf{c} and~\textbf{d} show the results for the pairs, which can be attributed either to local-distinguishable processes or the apparent teleportation. 
The plots~\textbf{a} and~\textbf{c} are obtained without smearing due to statistical hadronisation, whereas the plots~\textbf{b} and~\textbf{d} include the smearing. 
}
\label{fig:rap:Dmeson:200MeV}
\end{figure}

\begin{figure}
\centering
\includegraphics[width=.9\linewidth]{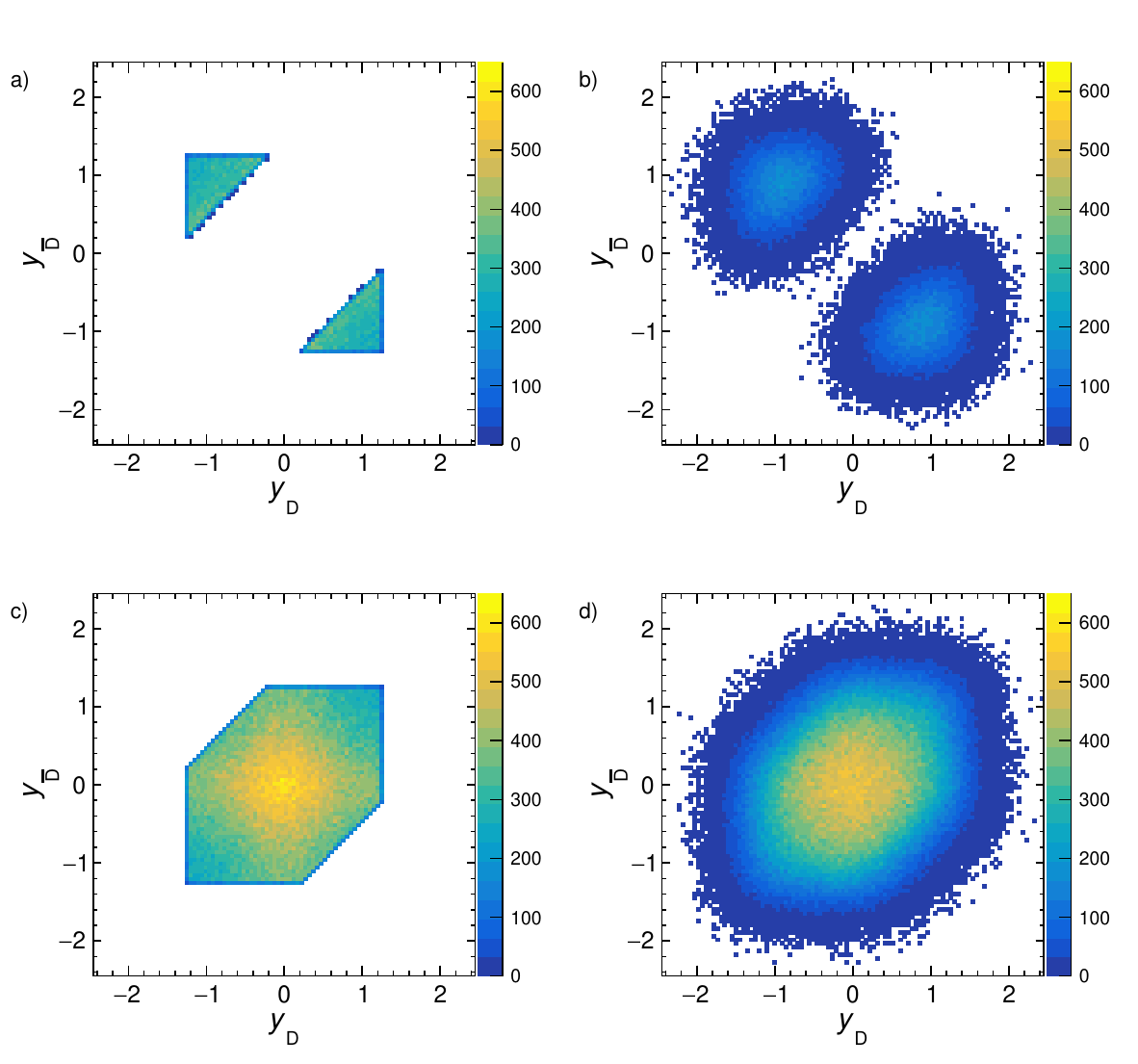}
\caption{
Joint rapidity distributions of $D$ and $\bar{D}$ mesons for the initial temperature $T_0 = 300$~MeV. 
The plots~\textbf{a} and~\textbf{b} show the distributions for $D$ and $\bar{D}$ pairs, which can result only from the apparent teleportation.
The~\textbf{c} and~\textbf{d} plots show the results for the pairs, which can be attributed either to local-distinguishable processes or the apparent teleportation. 
The plots ~\textbf{a} and~\textbf{c} were obtained without smearing due to statistical hadronisation, whereas the plots~\textbf{b} and~\textbf{d} include the smearing. 
}
\label{fig:rap:Dmeson:300MeV}
\end{figure}

Figures~\ref{fig:rap:Dmeson:200MeV} and~\ref{fig:rap:Dmeson:300MeV} show the 
joint rapidity ($y = tanh^{-1}(\beta)$) distributions of $D$ and $\bar{D}$ mesons for $\tau_{\mathrm{HAD}} = 10$~fm for $T_0 = 200$~MeV and $T_0 = 300$~MeV, respectively. The space for separating the apparent teleportation is significantly larger for the lower value of $T_0$. 
Including statistical hadronisation significantly smears the joint rapidity distributions. 

Our study shows that apparent teleportation can be observed in heavy-ion collisions at the SPS energy. The chance of success depends on the system's initial temperature and the hadronisation hypersurface.

\section{The 1+3D Blast-Wave model}
\label{app:B}

The 1+3D model of high-energy nucleus-nucleus collisions, with properties that allow for the prediction of experimentally measured quantities needed to observe the apparent teleportation, is presented here. The model utilises the Blast-Wave model~\cite{Schnedermann:1993ws} of radial flow, which has been widely used for parametrising experimental results on hadron spectra for decades. Here, only central Pb+Pb collisions are considered.

\noindent
The model assumptions are as follows.
\begin{enumerate}[(i)]
    \item The initial stage (\( t = 0 \)) of the matter created in a central Pb+Pb collision is approximated as a sphere of radius \( R = 6 \, \text{fm} \) (the radius of the Pb nucleus), centred at the origin of the coordinate system. The density is assumed to be isotropic.
    \item The hadronisation stage, where the radial flow stops, is approximated as a sphere with radius \( 2R \).
    \item The radial flow velocity at hadronisation at position \( \mathbf{r} = (x, y, z) \) is given by:
    \begin{equation}
    \label{eq:flow}
        \mathbf{v(r)} = c \cdot \mathbf{r} / (2R)~,
    \end{equation}
    where \( c \) is set to one. The hadronistation occurs at time \( t_{\text{HAD}} = R / v(r=2R) = c / R \).
\end{enumerate}

The distribution function of a pair of \( D \) and \( \bar{D} \) mesons generally depends on six momentum components (\( \mathbf{p}_D \) and \( \mathbf{p}_{\bar{D}} \)). However, due to the spherical symmetry of the model, this dependence reduces to three non-trivial momentum quantities. Here, we select them as:
\begin{itemize}
\renewcommand{\labelitemi}{-}
    \item The opening angle between the momentum vectors, \( \Theta \), where \( \Theta = |\theta_D - \theta_{\bar{D}}| \) and \( \theta_D \), \( \theta_{\bar{D}} \) are the polar angles of \( D \), \( \bar{D} \) mesons changing in the range \( [0, \pi] \).
    \item The momentum magnitudes \( p_D \) and \( p_{\bar{D}} \).
\end{itemize}

\begin{figure}[ht]
\centering
\includegraphics[width=0.8\textwidth]{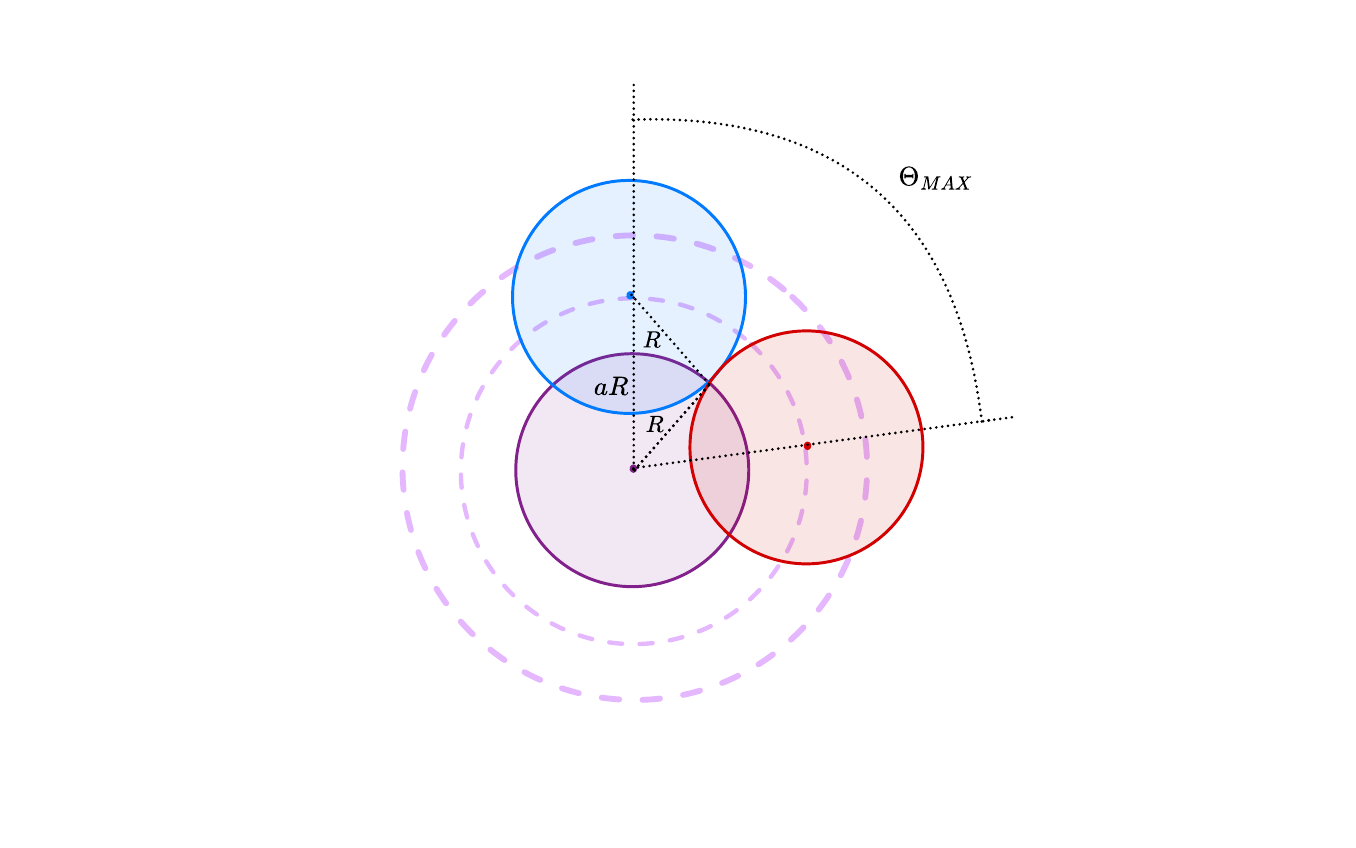}
\caption{
Sketch illustrating the calculation of the maximum possible opening angle \( \Theta_{\text{MAX}} \) between the \( c \) and \( \bar{c} \) quarks 
hadronising at \( r = a \cdot R \) and
obeying the LD condition. The plot shows the intersection of the plane defined by the hadronisation points and the collision centre with the past light-cone spheres of the \( c \) (red) and \( \bar{c} \) quarks, as well as the collision light cone at \( t = 0 \). The angle \( \Theta_{\text{MAX}} \) is determined by the requirement of a single common point of the three circles.
}
\label{fig:max_dist}
\end{figure}

\noindent
The model predictions are as follows.

Neglecting the non-flow component of the final velocity, if the hadronisation points of \( c \) and \( \bar{c} \) are chosen independently, the cosine of the opening angle, \( \cos(\Theta) \), is uniformly distributed. This set of pairs includes both local-distinguishable pairs and those violating this condition.

At \( t = 0 \), the collision light cone is a sphere with radius \( R \) centred at \( (0,0,0) \). The past light cones of the hadronising \( c \) and \( \bar{c} \) quarks on a sphere with radius \( r = aR \) (\( 0 \leq a \leq 2 \)) are also spheres of radius \( R \), but centred at \( \mathbf{r}_c \) and \( \mathbf{r}_{\bar{c}} \)
( \( r_c = r_{\bar{c}} = a R \) ), respectively. 

The maximum allowed opening angle \( \Theta_{\text{MAX}}(aR) \) for which the LD condition is obeyed is:
\begin{equation}
    \Theta_{\text{MAX}}(aR) = \pi - \cos^{-1}(1 - a^2 / 2)~.
\end{equation}
For \( a = 2 \) (the maximum hadronisation surface radius), \( \Theta_{\text{MAX}} = 0 \), while for \( a \to 1 \), \( \Theta_{\text{MAX}} \to 2/3~\pi \). For \( a = 1 \) and below, \( \Theta_{\text{MAX}} = \pi \). A sketch illustrating this calculation is shown in Fig.~\ref{fig:max_dist}.

The LD condition suppresses the opening angle distribution at large angles, with the effect strongest for pairs emitted from \( r = 2R \), where \( \Theta_{\text{MAX}} = 0^\circ \). For pairs hadronising at small radii (\( r \leq R \)) one gets \( \Theta_{\text{MAX}} = \pi \).

One can focus on pairs emitted from the outer hadronisation layer (high-momentum) to test for apparent teleportation. Observing pairs with \( \Theta > \Theta_{\text{MAX}} \) at high momentum would indicate apparent teleportation of \( c \) and \( \bar{c} \) quarks.

Various processes, in particular, hadronisation of (anti) charm quarks, will smear the opening angle distribution between 
\( D \) and \( \bar{D} \) mesons. 
Thus, separating the apparent teleportation will require correcting the opening angle distribution for the smearing
as discussed in Ref.~\cite{Gazdzicki:2023niq}.
As the smearing will be mostly due to effects that affect the emission angles of \( D \) and \( \bar{D} \) independently, the correction is expected to be approximately model-independent.

\acknowledgments
We thank F. Giacosa, M. Gorenstein, D. Miskowiec, and St. Mrowczynski for their comments.
This work is partially supported by
the Polish National Science Centre grants 2018/30/A/ST2/00226, 2018/30/E/ST2/00089 and 2020/39/D/ST2/02054.


\bibliographystyle{JHEP}
\bibliography{biblio.bib}

\end{document}